\documentclass[journal=aamick,manuscript=article]{achemso}
\usepackage{siunitx}
\usepackage{physics}
\usepackage[capitalize]{cleveref}
\usepackage{graphicx}
\makeatletter
\let\saved@includegraphics\includegraphics
\AtBeginDocument{\let\includegraphics\saved@includegraphics}
\renewenvironment*{figure}{\@float{figure}}{\end@float}
\makeatother
\usepackage{color}
\usepackage{xr}

\usepackage{amsmath}
\usepackage{soul}

\title[Ohmic contact engineering in few-layer black Phosphorus field effect transistors]
  {Ohmic contact engineering in few-layer black Phosphorus field effect transistors}
  
\author{Francesca Telesio}
\affiliation{NEST, Istituto Nanoscienze-CNR and Scuola Normale Superiore, Piazza San Silvestro 12, 56127 Pisa, Italy}
\email{francesca.telesio@nano.cnr.it}

\author{Gwenael le Gal}
\affiliation{NEST, Istituto Nanoscienze-CNR and Scuola Normale Superiore, Piazza San Silvestro 12, 56127 Pisa, Italy}

\author{Manuel Serrano--Ruiz}
\affiliation{Istituto di Chimica dei Composti Organometallici (CNR-ICCOM), Via Madonna del Piano 10, 50019 Sesto Fiorentino, Italy}

\author{Federico Prescimone}
\affiliation{Istituto per lo Studio dei Materiali Nanostrutturati-CNR, Via Piero Gobetti, Bologna, Italy}

\author{Stefano Toffanin}
\affiliation{Istituto per lo Studio dei Materiali Nanostrutturati-CNR, Via Piero Gobetti, Bologna, Italy}

\author{Maurizio Peruzzini}
\affiliation{Istituto di Chimica dei Composti Organometallici (CNR-ICCOM), Via Madonna del Piano 10, 50019 Sesto Fiorentino, Italy}

\author{Stefan Heun}
\affiliation{NEST, Istituto Nanoscienze-CNR and Scuola Normale Superiore, Piazza San Silvestro 12, 56127 Pisa, Italy}

\begin{document}

\begin{abstract}
Achieving good quality Ohmic contacts to van der Waals materials is a challenge, since at the interface between metal and van der Waals material, different conditions can occur, ranging from the presence of a large energy barrier between the two materials to the metallization of the layered material below the contacts. In black phosphorus (bP), a further challenge is its high reactivity to oxygen and moisture, since the presence of uncontrolled oxidation can substantially change the behavior of the contacts.  In this study, we investigate the influence of the metal used for the contacts to bP against the variability between different flakes and different samples, using three of the most used metals as contacts: Chromium, Titanium, and Nickel. Using the transfer length method, from an analysis of ten devices, both at room temperature and at low temperature, Ni results to be the best metal for Ohmic contacts to bP, providing the lowest contact resistance and minimum scattering between different devices. Moreover, we investigate the gate dependence of the current-voltage characteristics of these devices. In the accumulation regime, we observe good linearity for all metals investigated. 
\end{abstract}


\section{Introduction}

Ohmic contact engineering for van der Waals materials represents a challenge in boosting the performance of any device based on a two-dimensional material. This opens new fields of investigation, since at the interface between a metal and a van der Waals semiconductor, many different conditions could apply, ranging from the presence of a thick tunneling barrier to metallization of the van der Waals material below the contacts, which could dramatically modify device performance \cite{Allain2015, Leonard2011}. With graphene, many strategies have been proposed, from  creating defects by oxygen plasma to improve the adhesion with the metal \cite{Choi2011, Song2013}, to the engineering of one-dimensional contacts at the edge of encapsulated flakes \cite{Wang2013a}. On the other hand, in transition metal dichalcogenides, a local phase transition between the semiconducting and the metallic phase was shown to greatly improve contact resistance \cite{Kappera2014, Kappera2014a, Cho2015, Jena2014}, as did bias annealing \cite{Giannazzo2016}. Moreover, a local p--type doping induced by an oxygen plasma on MoS$_2$  has been demonstrated to induce ambipolar behavior, \cite{Giannazzo2017a} and the optimization of the contacts on these materials is still improving  as demonstrated by the very recent achievement of ultraclean interfaces between transition metal dichalcogenides and In/Au contacts \cite{Wang2019}.

Black phosphorus (bP) is a layered semiconductor \cite{Morita1986}, which has been widely studied in the last few years because of the possibility to exfoliate it to very thin layers, even down to the monolayer, called phosphorene. Few-layer black phosphorus attracted great interest in the scientific community since its demonstration in 2014 \cite{Castellanos-Gomez2014b}, because of the direct band gap, tunable by layer number from 0.3~eV in the bulk to approximately 2~eV for the monolayer \cite{Das, Castellanos-Gomez2015}. This feature, together with a carrier mobility which reaches up to 45000~cm$^2$/(Vs) for encapsulated few-layer bP at cryogenic temperatures \cite{Long2016}, is very interesting among van der Waals materials, since it places bP between graphene with its semimetallic behavior and the high band gap transition metal dichalcogenides. Moreover, bP has a peculiar crystal structure, puckered-in plane, with a strong crystalline anisotropy between the armchair and the zig-zag direction. This morphological anisotropy in the plane induces a strong anisotropy of the optical and thermal transport properties, \cite{Xia2014a,Lee2015a} as well as of the electronic properties, observed by magnetotransport experiments \cite{Hemsworth2016, Telesio2018a}.  

The high reactivity of bP in humid air, which is much more relevant for thin layers \cite{Favron2015a}, has so far been the major drawback for device applications of this material. In particular, the combined effect of oxygen and water, also in very small concentrations, is very disruptive for bP \cite{Huang2016, Luo2016}. P atoms react with oxygen forming phosphorus oxides that are hygroscopic, thus the flakes gradually hydrolyze and dissolve \cite{Castellanos-Gomez2014b}. Moreover, this process is accelerated by light \cite{Favron2015a}, especially in the UV spectral region \cite{Ahmed2017}.

Concerning contact resistance in bP, theoretical efforts have been made in modeling both monolayer and multilayer bP in terms of work function alignment \cite{Cai2014,Gong2014} and by complementary band calculations with quantum transport simulations for monolayer bP \cite{Pan2016}.

In a metal-semiconductor interface, the work function is generally the most relevant parameter to take into account to evaluate current injection, since the most common interface configuration is the formation of a Schottky barrier between the metal and the semiconductor \cite{Sze}. In this case, the better the alignment of the work function of the metal with the valence band (for holes) or conduction band (for electrons) of the semiconductor, the more the contact will be Ohmic, with a linear current-voltage ($I-V$) characteristics. When the alignment is poor, the carriers have to overcome a potential barrier to be injected into the semiconductor. This process can happen by thermionic emission or by tunneling, depending on bias and temperature, and both mechanisms lead to a deviation from linearity in the $I-V$ curves (\textit{i.e.} non-Ohmic contacts), with a flat low-bias trend with almost no current and a steep current increase above a critical bias voltage value. In the Schottky-Mott approximation \cite{Sze}, the Schottky barrier height (SBH) for electrons ($\Phi_n$) is given as $\Phi_n =\Phi_m - \chi$, where $\Phi_m$ is the work function of the metal and $\chi$ the electron affinity of the semiconductor. The SBH for holes ($\Phi_p$) is defined as $\Phi_p = E_g - (\Phi_m - \chi)$, with $E_g$ the semiconductor band gap. Since the charge transport in bP is dominated by holes, because of its intrinsic p-doping \cite{Morita1986}, the more relevant SBH for bP is $\Phi_p$.

Despite the interest in this topic, experimentally, up to now, just few studies on the contact resistance to bP were reported \cite{Das, Avsar,Perello2015, Das, Jiang2018, Wang2016a}. Among them, a big focus is on the transition between unipolar and ambipolar behavior in bP \cite{Perello2015, Das}. Even if several metals were inspected, it is difficult to extract a consistent picture, given also the differences in sample fabrication techniques among the various studies, but most of all because of the high reactivity of bP, whose uncontrolled oxidation, hard to prevent, introduces a large uncertainty in the observed behavior. One approach to address this issue is to extract as much information as possible from the same device, for example the contact resistance for different flake thicknesses using a stepped flake,\cite{Perello2015} or by depositing contacts of two different metals on the same bP channel.\cite{Du2014a}

Our approach is the opposite, making several devices for each metal, since our aim is to investigate whether the influence of the metal chosen for the contacts is robust against inhomogeneities between different samples and against residual oxidation, which could possibly occur during fabrication.

Here, we examine three contact metals, Titanium, Chromium, and Nickel, with different work functions: $\Phi_{Ni} \approx 5.0$~eV \cite{Du2014a}, $\Phi_{Cr} \approx 4.5$~eV \cite{Lide2008}, and $\Phi_{Ti} \approx 4.3$~eV \cite{Allain2015}, which are thus expected to produce different band alignments with bP ($\chi_{bP} \approx 4.4$~eV).\cite{Feng2016}
The three metals used in this study were chosen not only for their different work functions but also for their technological relevance, since they have a good adhesion to SiO$_2$/Si, the most common substrate for device fabrication, and they are frequently used as sticking layer for gold electrodes. We perform a systematic study of contact resistance using the transfer length method, and we crosscheck our results for consistency with two--probe and four--probe electrical transport measurements. Moreover, we study the gate dependence of the source-drain current and of the resistance of bP flakes and discuss the possible influence of contact resistance.

\section{Results and Discussion}

\subsection{Device geometry and sample preparation}

\begin{figure}
  \includegraphics[width=\textwidth]{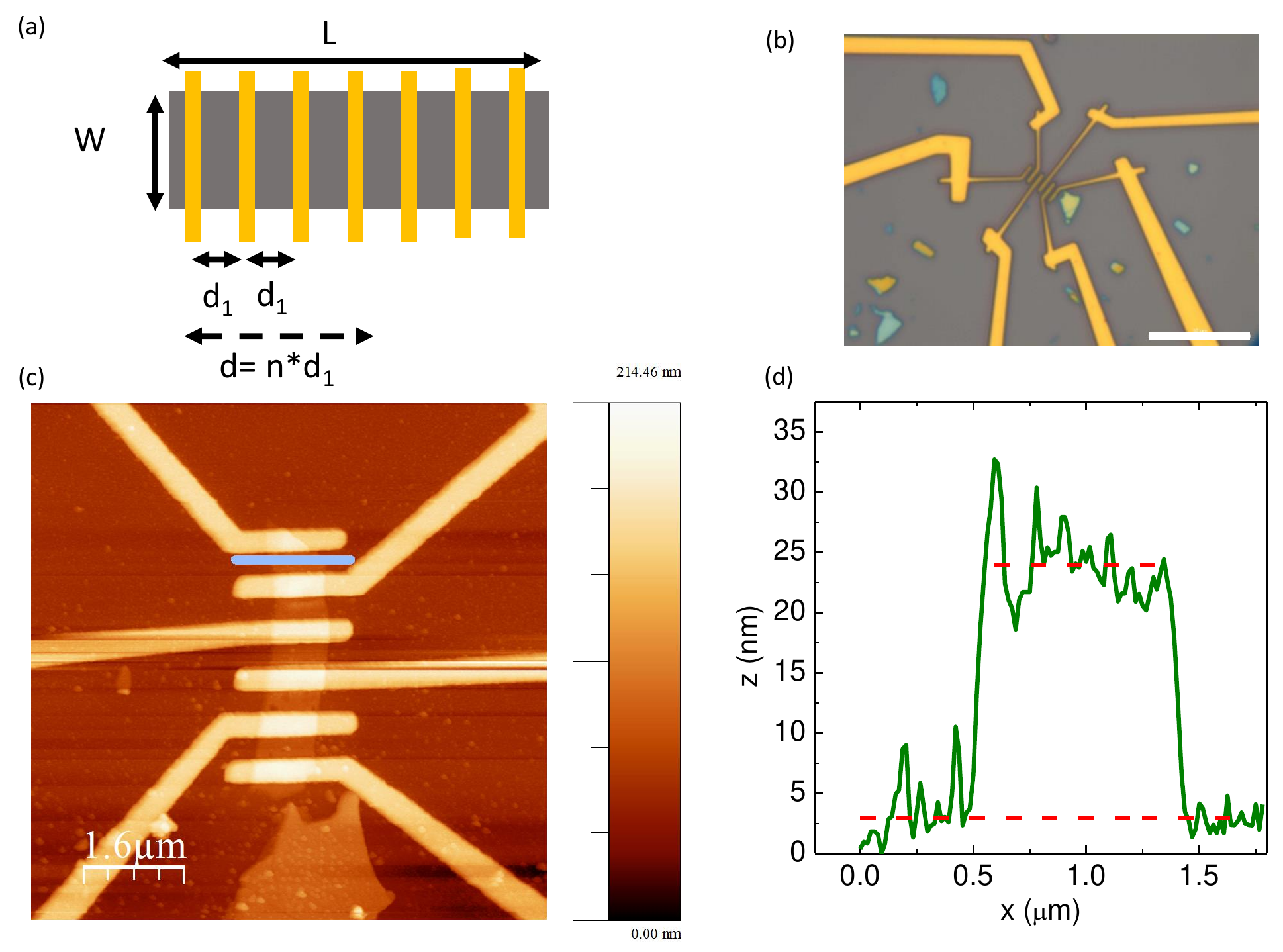}
  \caption{\label{fig:micro}(a) Schematics of the devices for transfer length method. Length ($L$) and width ($W$) of the bP flake are indicated. $d_1$ is the distance between neighboring contacts, while we label the general distance between two contacts as $d$. (b) Optical microscopy image of device 10. The scale bar is $10$~$\mu$m. (c) AFM image of the same device shown in (b). The scale bar is $1.6$~$\mu$m. (d) Profile along the line indicated in (c) to measure the flake thickness, which is $(21 \pm 4)$~nm. The red dotted lines indicate the average substrate and flake level.}
\end{figure}

The transfer length method (TLM) is one of the most frequently used methods to evaluate contact resistance \cite{Schroder}. It consists in measuring the two-terminal resistance ($R_2$) for various contact distances (Fig.~\ref{fig:micro}~(a)). The resistance measured in two--probe configuration is the series resistance of the two contacts and the channel. Assuming that all contacts between the metal and the bP flake of the device have the same resistance $R_C$, $R_2$ can be expressed as a function of the distance $d$ between the contacts, as $R_2 = 2 \cdot R_C + (d/W) \cdot R_S$, where $(d/W) \cdot R_S$ is the resistance of the bP channel, $R_S$ the sheet resistance of the flake, and $W$ its average width. We can thus extract $R_C$ and $R_S$ from a linear fit of $R_2$ vs.~$d$. In order to compare flakes of different width, we have to normalize the contact resistance $R_C$ to the flake width $W$, as shown in Fig.~\ref{fig:micro}(a).

In our device geometry, shown in Fig.~\ref{fig:micro}(a), we realize several ($n$) equidistant contacts at distance $d_1$. We measure $(n - 1)$ two--probe resistances from contacts at distance $d_1$, $(n - 2)$ two--probe resistances for contacts at distance $2 d_1$, and so on, up to a single measurement for the two outermost contacts.

Our aim is to investigate whether a general trend in contact resistance to bP is observed despite the scattering induced by possible residual oxidation and inhomogeneities. Therefore, the bP flakes used for this experiment, despite being from the same source, were selected from different crystals of different batches. Two different exfoliation protocols, one in a glove box and one in a glove bag, are used. bP is exfoliated on standard p-type Si wafers with 300~nm of SiO$_2$. The samples are then immediately coated with a bilayer composed of a methyl methacrylate methacryclic acid copolymer layer and a poly (methyl methacrylate) layer (~(MMA(8.5)MAA)/PMMA~) which acts both as a protective layer and as a resist for electron beam lithography (EBL). For the devices, elongated bP flakes were chosen, which showed a regular shape and an optical contrast typical of flakes of few tens of nanometers thickness. Several parallel contacts were designed on the flakes by EBL, at nominal distances $d_1$ ranging from 600~nm to 1~$\mu$m, depending on flake length $L$ (Fig.~\ref{fig:micro}(a)). Overall, 10 devices were realized and characterized, both at room temperature and at low temperature.

After the EBL step, we performed a mild oxygen plasma. This step allows to obtain more stable contacts for two reasons: first of all, it removes the organic residuals of the EBL resist after development, and furthermore, it introduces in a controlled way a thin oxide layer on the bP flake, which makes the samples more homogeneous and increases the quality of the interface.\cite{Szkopek2018} Oxygen plasma in bP treatment has indeed already found some applications, since phosphorus oxides have been observed to be a good dielectric when used in combination with alumina \cite{Szkopek2018}, and moreover, a controlled thinning of the flake can be achieved with a mild oxygen plasma \cite{Jia2015a}, as well. 

The metals are thermally evaporated onto the samples, which then undergo a lift-off process. Again a bilayer of (MMA(8.5)MAA)/PMMA is spun onto the chips to prevent degradation of the flakes during measurements. An optical microscopy image of a typical device is shown in Fig.~\ref{fig:micro}(b). Further details on the individual processes performed for device fabrication and measurements are given in the Methods section.

To measure the geometrical parameters of the flakes, after the electrical transport measurements, we removed the protective layer to perform atomic force microscopy (AFM) measurements. One example is shown in Fig.~\ref{fig:micro}(c). This process was performed as a last characterization step, since an AFM measurement in air degrades the flakes. These measurements allowed to precisely evaluate the width $W$ and length $L$ of the bP flakes and to measure the distances $d$ on each flake. The scattering in $d$ with respect to the nominal value resulted smaller than 5\% for all devices. Furthermore, we extracted flake thickness from cross-sections such as the one shown in Fig.~\ref{fig:micro}(d). The thickness of the individual flakes is in the range $15 - 40$~nm, which coincides with the thickness range found in literature on bP contacts \cite{Cai2014, Perello2015}. All flakes can be considered to be in the same thickness range for what concerns band alignment. The geometrical dimensions of the individual flakes are presented in the Supporting Information (Table~S1).

\subsection{Contact resistance}

The devices were measured at room temperature and at low temperature (4.2~K). As a standard for the analysis, we used a bias voltage in the $\pm 1$~mV range, which is the one typically used in experiments. In order to evaluate $R_2$, we acquired a full current--voltage ($I-V$) curve for each pair of contacts and extracted the two--probe resistances from linear fits of these curves. Examples of $I-V$ characteristics are shown in the Supporting Information in Fig.~S1. We found that the $I-V$ behavior of all devices is linear in this bias range, independent of contact metal. The linear behavior of the $I-V$ curves demonstrates the good quality of the contacts. Furthermore, it is crucial for the correct evaluation of the two--probe resistance $R_2$.

\begin{figure}
  \includegraphics[width=\textwidth]{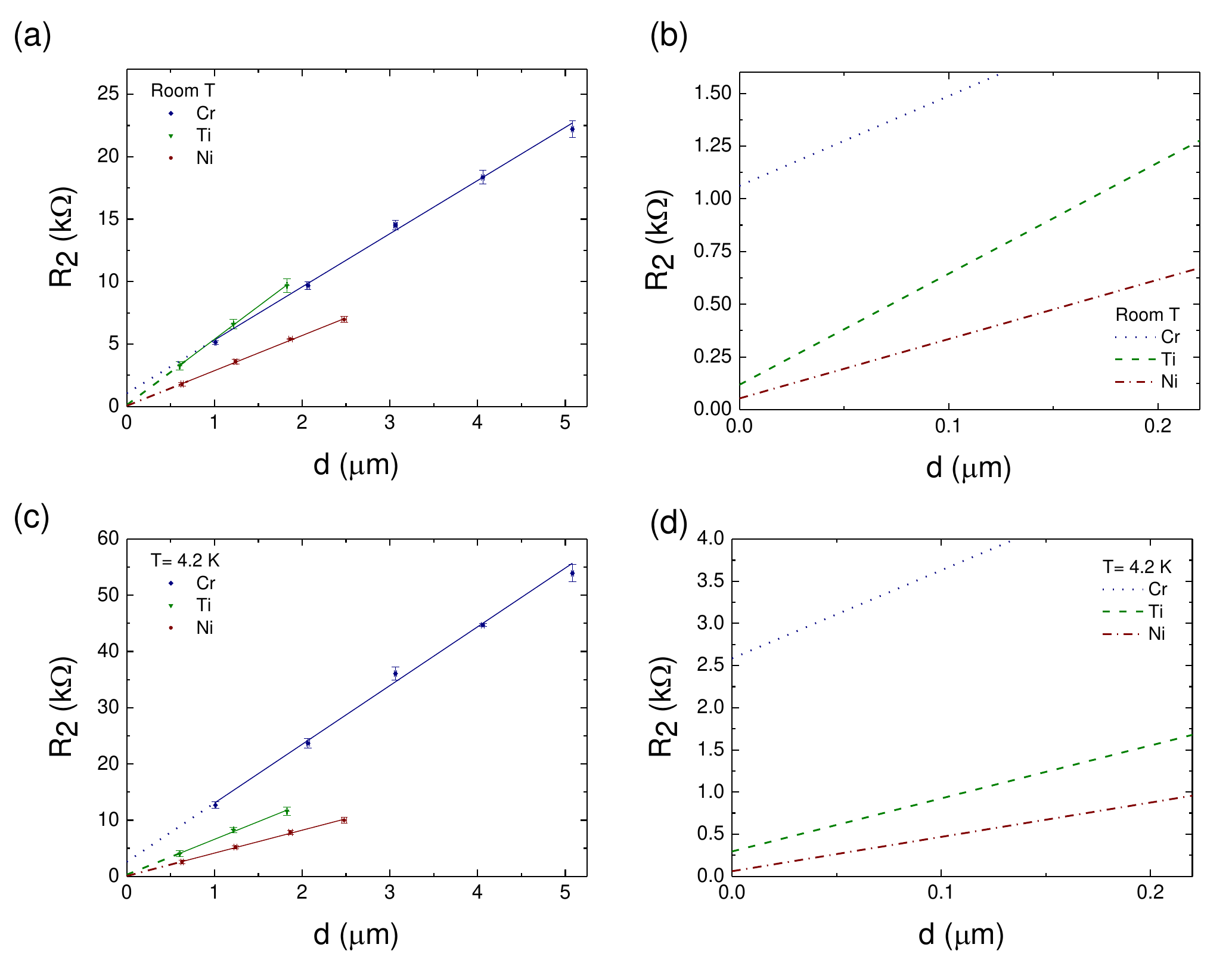}
  \caption{\label{fig:meas}Transfer length method: average two--probe resistance $R_2$ at room temperature (a) and at low temperature (c) as a function of contact distance $d$. The solid lines in (a) and (c) are the results of the linear fit procedure. Outside the range of fitting, the same linear function is plotted using dashed lines. A zoom--in showing the intercepts at $d=0$, corresponding to $2R_C$, is shown in (b) and (d) for room temperature and low temperature, respectively. Ni: device 7, Ti: device 5, Cr: device 2.}
\end{figure}

Plots of $R_2$ vs.~$d$ for three representative devices, one for each metal, are presented together with linear fits in Fig.~\ref{fig:meas} both for room temperature (panels (a,b)) and low temperature (panels (c,d)). The linear trend expected by TLM is indeed observed, and the scattering of the resistance values for the same distance $d$ is very small, as visible from the error bars, which are often barely visible in the graph. From the slope of the linear fits, the sheet resistance $R_S$ is obtained, while the intercept is equal to $2R_C$. To better inspect this region of interest, a zoom-in at $d$ close to 0~$\mu$m is shown in Figs.~\ref{fig:meas}(b,d).

We then calculated $R_C \cdot W$ and averaged the results ($\overline{R_C \cdot W}$) for all devices which had contacts of the same metal, in order to rule out as much as possible device-dependent effects and to obtain the general trend related just to the metal bP layer contact. These results are presented in Table~\ref{RCtab1}. Two different errors are evaluated on the aggregate data: the propagated error on the average, labeled as avg.~error, which is related to the experimental errors during the measurements and the scattering of the data on the single device, see Fig.~\ref{fig:meas}. The other relevant uncertainty is the standard error, labeled in the table as std.~error, which is the standard deviation of the distribution of contact resistances of different devices with the same contact metal divided by $\sqrt{N}$, where $N$ is the number of devices. This latter value allows an evaluation of the scattering between different devices with the same metal. Further information on the evaluation of uncertainties is presented in Section 2 of the Supplementary Information.

Figure~\ref{fig:meas} shows that contact resistance at low temperature is higher than at room temperature, which suggests a contribution of bP to $R_C$ for all three metals. This trend is also clearly visible from the contact resistance data presented in Table~\ref{RCtab1}. At room temperature, $\overline{R_C \cdot W} = 135$ ~$\Omega \: \mu$m is obtained for Ni, which is among the lowest previously reported contact resistances in the literature for bP\cite{Ma2017,Park2018,Huang2019} and well within the 2017 target for silicon transistors in the International Technology Roadmap for Semiconductors (ITRS).\cite{Liu2016} Ti has a normalized contact resistance which is slightly higher ($\sim 200 \: \Omega \: \mu m$), but still much lower than values reported in literature for this metal,\cite{Haratipour2017} while the highest result is obtained for Cr with $\overline{R_C \cdot W}$(TLM) $\sim 800$~$\Omega \: \mu$m. Also at low temperature, chromium with $\sim 2400$~$\Omega \: \mu$m is the metal that shows the highest contact resistance, followed by Ti with 740~$\Omega \: \mu$m. Ni with 430~$\Omega \: \mu$m confirms to be the metal with the best performance, and also in this case $\overline{R_C \cdot W}$ of Ni is more than a factor five smaller than for Cr. Moreover, with the lowest scattering of the results, Ni contacts show the most consistent performance among the three metals investigated.

To obtain another independent evaluation of contact resistance, we compared four--probe and two--probe measurements ($R_2 \: vs \: R_4$). For this purpose, the outermost contacts were used as source and drain, while all other contact combinations were used for the four--probe resistance measurements ($R_4$). Four--probe resistance measures just the contribution of the bP channel, $R_4 = R_S \cdot d/W$. We compare this value with the corresponding two--probe resistance $R_2 = 2 \cdot R_C + R_S \cdot d/W = 2 \cdot R_C + R_4$, measured using the same \textit{inner} contacts as for the four--probe measurement. We can thus evaluate $R_C$ as $R_C=(R_2-R_4)/2$. This method allows to eliminate the influence of any possible inhomogeneity of the bP channel, because the very same region of the channel is probed in both measurements. In these measurements, which are summarized in Table~\ref{RCtab1}, Ni shows again the best performance, with specific contact resistivity values which are consistent with those obtained by TLM. Also for Ti, a good agreement between the two methods is obtained, especially at low temperature, while for Cr comparatively large variations are observed.

\begin{table}
\center
\begin{tabular}{|c|c||c|c|c||c|c|c|} \hline
& & \multicolumn{3}{c||}{Room T} & \multicolumn{3}{c|}{Low T (4.2 K)} \\ \hline
metal & method& $\overline{R_C \cdot W}$ & avg.~error & std.~error & $\overline{R_C \cdot W}$ & avg.~error & std.~error  \\
& & $\Omega \: \mu$m & $\Omega \: \mu$m & $\Omega \: \mu$m & $\Omega \: \mu$ & $\Omega \: \mu$m & $\Omega \: \mu$m \\ \hline
Cr & TLM & 797 & 64 & 253 & 2428 & 198 & 1377 \\ 
& $R_2 \: vs \: R_4$ & 444 & 165 & 0.2 & 1353 & 240 & 408 \\ \hline
Ti & TLM & 217 & 88 & 163 & 740 & 209 & 282 \\ 
& $R_2 \: vs \: R_4$ & 484 & 56 & -* & 912 & 225 & 382 \\ \hline
Ni & TLM & 135 & 13 & 64 & 432 & 30 & 208   \\ 
& $R_2 \: vs \: R_4$ & 138 & 25 & 75 & 448 & 64 & 213   \\ \hline
\end{tabular}
\caption{\label{RCtab1}Average of the normalized contact resistance $\overline{R_C \cdot W}$ obtained from the transfer length method (TLM) measurements and from the comparison between two--probe and four--probe resistances ($R_2 \: vs \: R_4$), both at room temperature and at 4.2 K, for each of the three metals investigated. Both the propagated error on the average (avg.~error) and the standard error (std.~error) are displayed. *: The standard error is missing because these measurements are from a single device.}
\end{table}

Table \ref{RCtab1} shows that the standard error is typically larger than the average error. This is expected because the standard error describes the scattering between the different devices and includes therefore the variation of the characteristics of the specific flakes, which were either intentionally allowed to vary (flake thickness or processing) or will otherwise have an influence on contact resistance, for instance the crystallographic orientation of the flakes (which we did not control).

Among the metals investigated, in the Schottky-Mott model only Ni has a negative energy barrier for hole conduction ($\Phi_{p,Ni} = -0.3$~eV). On the other hand, for the case of Cr and Ti, formation of a barrier is predicted by the model, with $\Phi_{p,Cr} = 0.2$~eV and $\Phi_{p,Ti} = 0.4$~eV. Thus, Ti has the worst work function alignment \cite{Allain2015, Cai2014}. Still, the performance of the Cr contacts is more moderate, both in terms of contact resistance and for what concerns the scattering from device to device. Therefore, the Schottky-Mott rule does not capture the full physics occurring at the interface. However, it is well known that Fermi level pinning \cite{Schroder} can occur, which leads to a deviation from the ideal trend.\cite{Wang2018} The presence of Fermi level pinning has already been suggested and studied theoretically for few--layer bP \cite{Lee2017b}, and it has been as well investigated experimentally \cite{Jiang2018}. What is observed for bP \cite{Jiang2018} (and what is generally observed in semiconductors) is a combination of Schottky--Mott barrier and Fermi level pinning. The configuration of the surface states in the semiconductor is very important in determining Fermi level pinning.\cite{Bardeen1947, Tung2014} Since in our sample processing we introduced an oxygen plasma step, this could have modified the surface state density, changing the conditions at the interface. The same process might even have created a thin insulating tunnelling barrier of $PO_x$ which could weaken the band matching condition. In addition to this, inhomogeneities at the interface can play a crucial role for Fermi level pinning \cite{Tung1993a,Tung2014}. A theoretical prediction suggests a more defected growth of Cr on bP surfaces with respect to Ni. From energetic considerations, it appears that a three--dimensional island growth of Cr on bP is favourable, while for Ni a two--dimensional growth is predicted. \cite{Hu2015b}. This could be relevant, since defects, interface dipole moments, and the chemistry at the interface can produce significant deviations from the simple Schottky-Mott model \cite{Tung2014}. Moreover, independent theoretical studies on monolayer bP suggest that at the interface with Cr, non--trivial effects occur, leading to a better transport for electrons than for holes\cite{Pan2016}.

Table~\ref{AGGRnmu} gives the values of sheet resistance $R_S$ obtained from the slope of the linear fits in Fig.~\ref{fig:meas}. Besides, we obtain an independent estimate of sheet resistance from the four--probe measurements, since $R_S = R_4 \cdot W / d$. Data obtained with TLM and $R_2vR_4$ methods display both at room and low temperature a very good consistency for all three metals investigated. At room temperature, devices from all three metals have similar sheet resistance. At low temperature, instead, the sheet resistance for Cr--contacted samples increases much more than for the other two metals, reaching 18~k$\Omega$, which is further evidence for an interfacial reaction between Cr and bP.

\begin{table}
\center
\begin{tabular}{|c|c||c||c|} \hline
metal & method &  Room T, $\overline{R_s}$ [k$\Omega $ ] &   Low T (4.2 K), $\overline{R_s}$ [k$\Omega$]    \\ \hline
Cr & TLM & $7.5\pm1.4$ &  $17.3\pm 8.0$  \\ 
 & $R_2 \: vs \: R_4$ & $6.5\pm0.7$ &    $18.0\pm9.0$    \\ \hline
Ti & TLM & $7.0\pm1.5$  &  $ 9.3\pm2.0$    \\ 
& $R_2  \: vs \: R_4$ & $7.0\pm1.4$&    $9.8 \pm 2.6$     \\ \hline
Ni & TLM & $4.9\pm1.5$ &  $5.8\pm1.0$      \\ 
 & $R_2 \: vs \: R_4$ & $4.8\pm1.5$      &$5.7\pm0.9$         \\ \hline
\end{tabular}
\caption{\label{AGGRnmu} Average sheet resistance for all three metals under study. Only the larger between avg.~error and std.~error is shown in the table for clarity.}
\end{table}

To summarize, Ni provides the best contacts to bP, not only for its lower $\overline{R_C \cdot W}$ both at room and at low temperature, but also for the lower average error and standard error of the Ni contacts, and for the consistency between TLM and $R_2 \: vs \: R_4$. This underlines the good homogeneity of Ni contacts to bP. Moreover, the contact resistances obtained compare favorably to what is observed for bP and other 2D semiconductors\cite{Liu2018} and are not far from the quantum limit.\cite{Jena2014}

\subsection{Gate dependence}

\begin{figure}
  \center
  \includegraphics[width=0.5\textwidth]{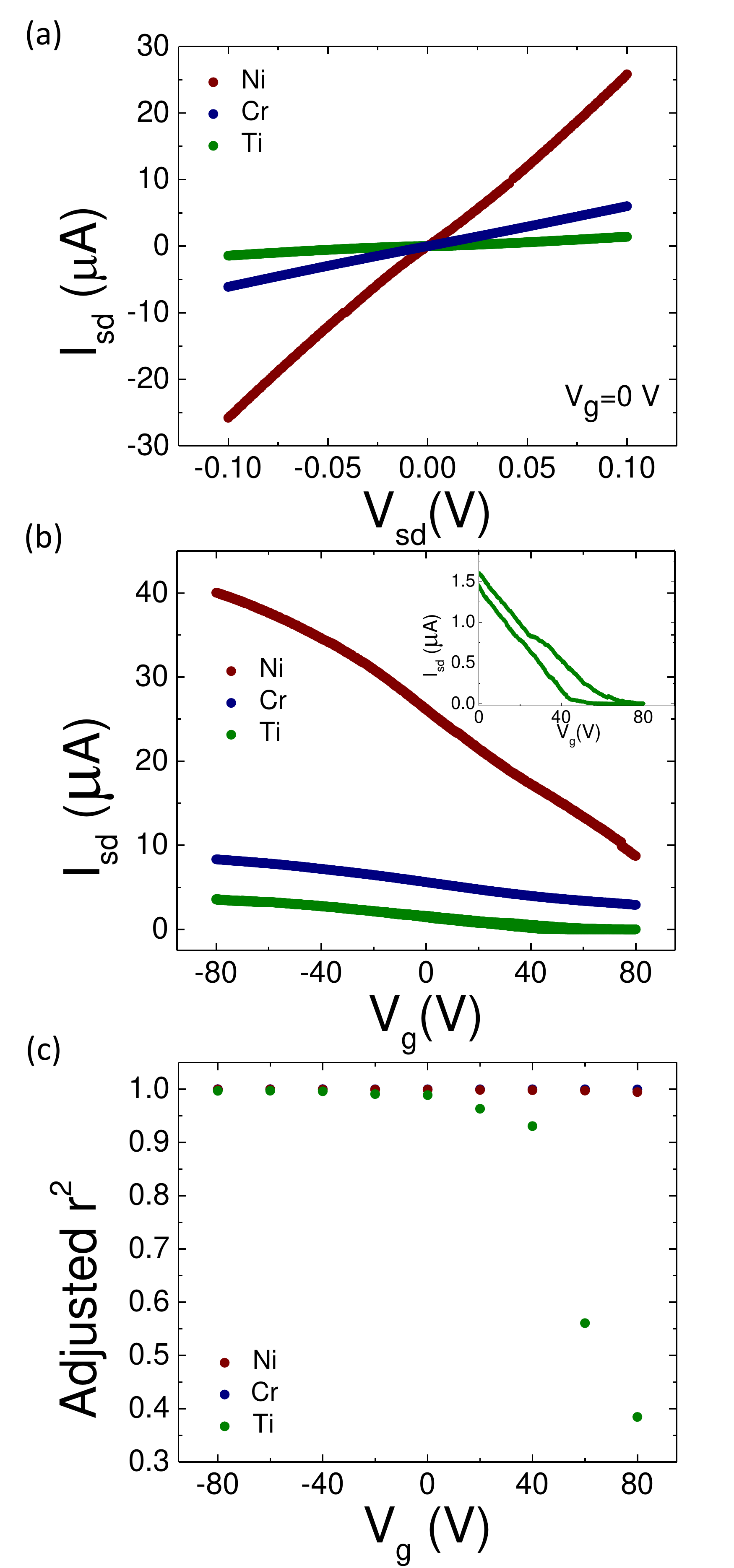}
  \caption{\label{fig:highV}(a) Two--probe current--voltage characteristics in the $\pm 100$~mV bias range for three specific devices, one for each of the selected metals (Ni, red curve, device 7; Cr, blue curve, device 3; and Ti, green curve, device 6). (b) Current versus gate voltage for the same devices. $V_{sd} = 100$~mV. The inset shows a zoom of the measurement relative to the Ti--contacted device, which displays the presence of a hysteresis in the depletion regime.  (c) Adjusted $r^2$ of the linear fits of all current-voltage measurements acquired at different gate voltages. All  measurements taken at $T = 4.2$~K.}
\end{figure}

Generally the $\pm 1$~mV range is suitable for performing transport measurements, since higher voltages could produce self heating in the proximity of contacts  due to current crowding \cite{Wang2016a}. Nevertheless, at low temperature we inspected a wider bias voltage range, namely $\pm 100$~mV, to search for nonlinearities, which are an indication of a Schottky contact to bP. A typical $I-V$ curve for a device of each metal is shown in Fig.~\ref{fig:highV}(a). To quantitatively evaluate the linearity of the $I-V$ curves, we performed a linear fit of each curve and used the adjusted $r^2$ of the fit as an indicator of the linearity. As can be seen in Fig.~\ref{fig:highV}(a), with no gate bias applied ($V_g = 0$~V), all curves are linear, with an adjusted $r^2 \approx 0.99$ or higher.

Source--drain current vs.~gate voltage measurements are shown in Fig.~\ref{fig:highV}(b). These measurements clearly show that all samples display a strong unipolar p-type behavior, independent of the contact metal used. Several factors could contribute to this effect, such as flake thickness, since we realized our FETs with bP multilayers in the 15~nm to 40~nm thickness range, or the 300~nm thickness of the SiO$_2$ used as a gate dielectric, which could be too thick to efficiently gate bP into the electron region \cite{Das, Du2014a}. Another factor which could play a role is the oxygen plasma procedure that we use, since, as suggested by some preliminary evidence,\cite{Perello2015} oxygen exposure could increase the p--type doping of bP. However, a systematic investigation of the link between oxygen exposure and p--type doping is still missing.

Figure~\ref{fig:highV}(b) shows that for positive gate voltage above 40~V, one device (with Ti contacts) reaches the depletion regime, highlighted with a more appropriate scale in the inset to Fig.~\ref{fig:highV}(b), and shows a hysteresis there. The hysteresis in bP FETs on SiO$_2$ has been previously observed in literature  and related to trap states at the interface \cite{Illarionov2016}. A similar behavior was found for another device (with Ni contacts), which underlines that this effect is not related to the contact metal type. Indeed, measurements provided in the Supplementary Information (Fig.~S2) clearly show that only when the depletion regime is reached, the $I-V$ characteristics in these devices becomes non--linear. 

The adjusted $r^2$ of linear fits to current-voltage measurements at various gate voltages are displayed in Fig.~\ref{fig:highV}(c). While in the accumulation regime all contacts show a linear trend (adjusted $r^2 \approx 1$), the adjusted $r^2$ drops when a device reaches the depletion regime, here $r^2 < 0.8$ for $V_g > 40$~V for the device with Ti contacts. A similar deterioration of contact quality in the depletion regime has already been observed in bP \cite{Li2014}.

\section{Conclusions}

In this work, we presented a study of contact resistance in exfoliated black phosphorus devices with thickness $15 - 40$~nm, in a transfer length method configuration. Three different metals, Titanium, Chromium, and Nickel were used, and measurements were performed both at room temperature and at liquid Helium temperature. In order to evaluate how robust the observed trends are, we presented aggregate data from 10 devices and used two different exfoliation protocols for bP.

Both at room temperature and at low temperature, contacts from all three metals show linear $I-V$ curves in the $\pm 1$~mV regime. Devices contacted with Cr display the highest contact resistances and the largest scattering of the data. The anomalous behavior of Cr--contacted devices can be related to a more defective growth mechanism of Cr on bP, consistent with theoretical studies \cite{Hu2015b}. The best contacts are achieved with Ni. Thus, Nickel is the best performing contact metal for low temperature electrical transport experiments with bP. This trend is confirmed by the comparison between two--probe and four--probe resistance measurements.

In all samples we observed  a strong unipolar p-type behavior. We observe a linear behavior for all three metals in the $\pm 100$~mV bias voltage regime, when the samples are not in the depletion regime. Non-linearities appear just when the depletion regime is reached and $I_{sd}$ approaches zero, as previously already observed in bP FETs \cite{Li2014}.

\section{Methods}

\subsection{Materials}

In our experiments, we used bP crystals prepared according to a published procedure \cite{Kopf2014}, wherein high-purity red phosphorus ($>99.99$\%), tin ($>99.999$\%), and gold ($>99.99$\%) are heated in a muffle oven with a SnI$_4$ catalyst. The solid product was placed in a quartz tube, subjected to several evacuation-purge cycles with N$_2$ gas, and then sealed under vacuum. The evacuated quartz tube was heated at 4.2~$^{\circ}$C/min to 406~$^{\circ}$C, where it remained for 2~hours. The tube was then heated at 2.2~$^{\circ}$C/min to 650~$^{\circ}$C and held at this temperature for 3 days. Finally, the tube was cooled to room temperature at 0.1~$^{\circ}$C/min. The product is crystalline bP with a typical size of some mm.

\subsection{Device fabrication}

Substrates of Boron-doped Si, with 300~nm thermal SiO$_2$, were pre-patterned by optical lithography with a grid of markers to allow flakes identification. Bonding pads were included in this lithography step. For device fabrication, the commercial MMA(8.5)MAA copolymer (EL13) and PMMA (A4), both from MicroChem, were used as a positive  resist for electron beam lithography. EBL was performed with a Zeiss UltraPlus Scanning Electron Microscope (SEM), equipped with an interferometric stage for better alignment, and with a Raith tool for EBL. An acceleration voltage of 20~kV and a dose of 350~$\mu$C/cm$^2$ were used. Two different apertures were employed, namely 10~$\mu$m and 30~$\mu$m/60~$\mu$m, corresponding to currents of 33~pA and 190~pA/560~pA, for the active area of the device and for the gold paths to the bonding pads, respectively. Having a lower current for the exposure of the active area of the device allowed to have a precise patterning of the device, while the part exposed with higher current was optimized for fast exposure time. The development was performed in AR 600-56, a commercial developer from Allresist GmbH, based on 4-Methylpentan-2-one diluted in IPA. 

Oxygen plasma was performed in a Sistec reactive ion etching system, with a flux of 40 standard cubic centimeters per minute (sccm) and a power of 10~W for 1~minute, after a cycle of chamber cleaning to reduce at best any contamination.

Metal evaporation was performed in a Sistec multi-crucible thermal evaporator equipped with a rotating carousel, in order to have the sample aligned with the metal crucible. We evaporated 10~nm of the selected metal, with an evaporation rate of  0.4~\AA /s for Nickel, 0.3~\AA /s for Chromium, and 3~\AA /s for Titanium. An overlayer of 100~nm gold was then evaporated at 1.5~\AA /s. The pressure in the evaporator chamber was around $10^{-6}$~mbar before evaporation, and  increased slightly during metal evaporation. After metal evaporation the samples underwent a 15 minutes lift-off in hot acetone.

The same two resists used for fabrication were used as protective layer, to avoid device aging between measurements and, through a second electron beam lithography step, holes were opened on top of the bonding pads, to connect the samples with the chip carriers on which they were mounted.

\subsection{Measurement setup}

For electrical transport measurements, the samples were bonded to a 16 pin dual in line chip carrier using Au wire. The transport properties were measured in DC or AC under vacuum ($p < 10^{-4}$~mbar) in a custom--made insert, equipped with a diode for temperature measurement. The leakage current though the gate was measured during $V_g$ loops and always found to be negligible.

DC measurements were mainly used for current-voltage measurements and two--probe gate sweeps. These measurements were performed using a DC voltage source with a slow sweep rate, while the current was measured with a DL current preamplifier and an Agilent 34410A digital multimeter. Four--probe measurements were mainly performed in AC using Stanford Research SR830 lock-in amplifiers, at a frequency of 17~Hz with a 10~M$\Omega$ resistance in series to the sample.

Atomic force microscopy (AFM) measurements were performed with a Bruker Dimension Icon AFM, in peak force mode. Data  analysis was performed with the WSxM software \cite{Horcas2007}.

\begin{acknowledgement}

The authors would like to thank Alessandra Campana for her support during the initial stage of this work. This work was financially supported by EC through the project PHOSFUN {\it Phosphorene functionalization: a new platform for advanced multifunctional materials} (ERC ADVANCED GRANT No. 670173 to M.~P.). S.~H. thanks Scuola Normale Superiore for support, project SNS16\_B\_HEUN -- 004155. F.~T. thanks CNR-Istituto Nanoscienze for funding the SEED project 2017 SURPHOS. 
\end{acknowledgement}

\pagebreak

\section*{Supplementary Information to Ohmic contact engineering in few-layer black Phosphorus field effect transistors} 

\renewcommand{\thesection}{S\arabic{section}}
\renewcommand{\thefigure}{S\arabic{figure}}
\renewcommand{\thetable}{S\arabic{table}}
\setcounter{figure}{0}
\setcounter{table}{0}
\setcounter{section}{0}

\section{Current versus Voltage Measurements}

\begin{figure}[!h]
  \includegraphics[scale=0.55]{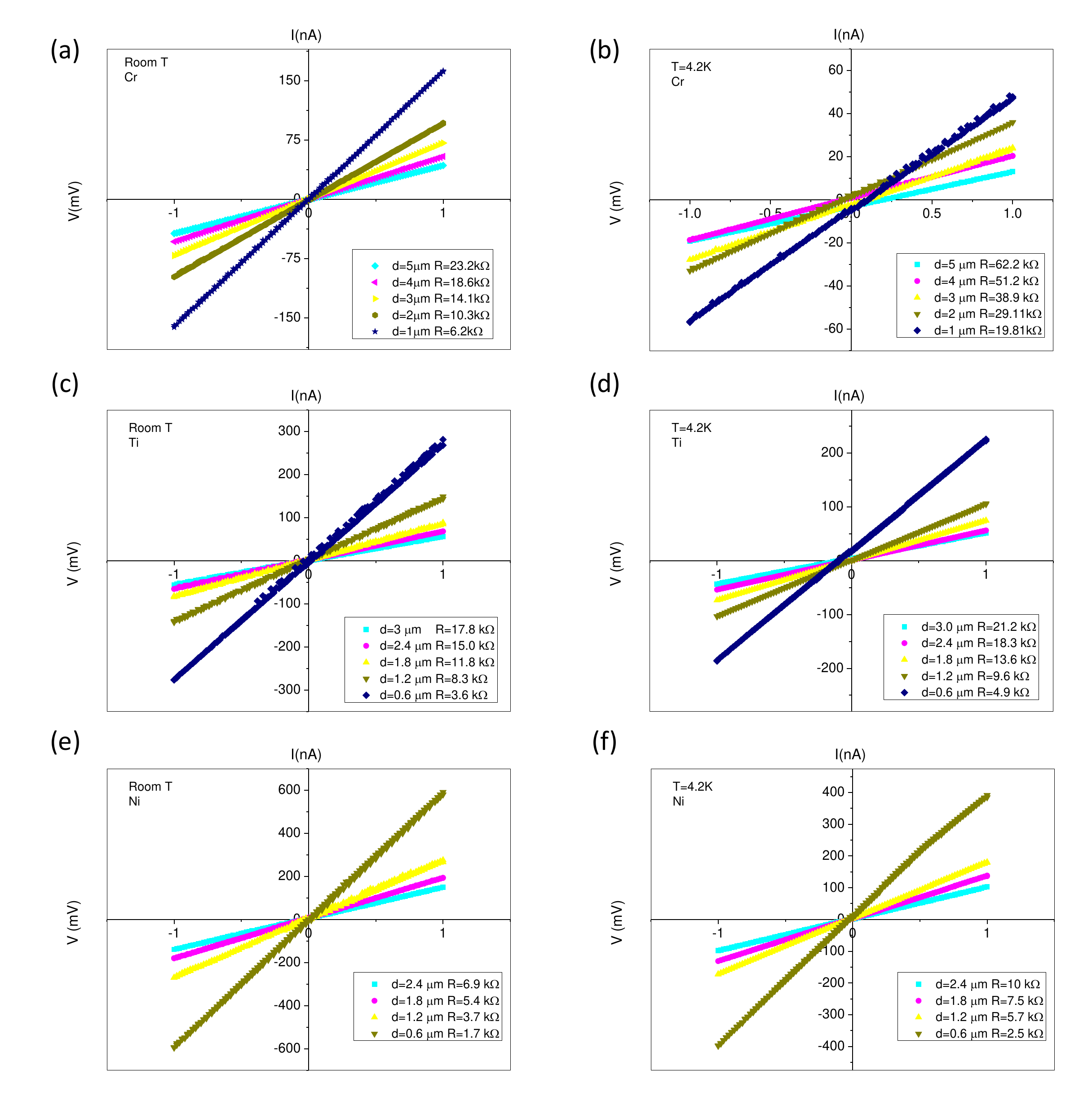}
  \caption{\label{fig:iv}Examples of current-voltage sweeps at room temperature (left column) and at low temperature (right column), for contacts at different distances, for each of the three metals investigated. Top row: Cr, panels (a,b), device~2; center row: Ti, panels (c,d), device~5; bottom row: Ni, panels (e,f), device~7.}
\end{figure}

\section{Evaluation of Uncertainties}

Since we want to evaluate how robust the contact resistance is against several variability factors, a careful analysis of the scattering of our data has to be made, to determine if and where a scattering is introduced in the results. After acquiring the raw $R_2$ data for all combinations of contacts present on each device, the two--probe resistances of all pairs of contacts with same $d$ are averaged, and the standard error is calculated. A standard deviation equal to the average of all standard deviations obtained for the same device is attributed to those contacts for which only a single measurement is available. These error bars are the ones reported in the two--probe resistance versus distance plots in Fig.~2 of the main text, from which we obtain the individual contact resistance values.

Since we have more than one device for each metal under investigation, in order to extract the aggregate data for each metal, we evaluate two different errors on the average, related to different uncertainty sources.

First of all, we propagate the errors $\Delta (R_C \cdot W)$ on the individual $R_C \cdot W$ values, as $\frac{1}{N} \sqrt{\sum (\Delta(R_C \cdot W))^2}$, with $N$ the total number of devices made with that metal. This error, that we call average error (avg.~error), provides information on the uncertainty of $R_C \cdot W$ in the single devices, but it does not contain information on the scattering between different devices.

To evaluate the scattering between the different devices with the same contact metal, we calculate the standard error (std.~error) on the distribution as $\frac{1}{\sqrt{N}} \sqrt{\sum (R_C \cdot W - \overline{R_C \cdot W})^2}$, where $\overline{R_C \cdot W}$ is the average normalized contact resistance for devices made using contacts of the same metal. We give the standard error of the distribution and not the standard deviation for a better comparison with the average error.

Similarly, in the evaluation of $R_C$ from the comparison between two--probe and four--probe measurements, discussed in the main text, we first evaluate the scattering of the data in the single device, and then we calculate $R_C \cdot  W$, its average error and its standard error.

For the sheet resistance $R_S$, we evaluate the errors using the same procedure. 

For the evaluation of  $R_C$, $R_C \cdot W$ and $R_S$, the dimensions of the flake and of the contacts play a role: width and length of the flakes, as well as contacts distances are obtained from AFM images.

\section{Geometrical Dimensions of the Devices}

\begin{table}[h]
\center
\begin{tabular}{|c|c|c|c|c|c|c|} \hline
device & metal & thickness $t$ & width $W$ & number of   & $d_1$   \\ 
\# & & [nm] & [$\mu$m] &  contacts & [$\mu$m]  \\ \hline
1 & Cr & $19\pm3$ & $1.63\pm0.09$ &4& $1.021\pm0.017$  \\ \hline
2 & Cr & $40\pm3$ & $1.58\pm0.11$ & 5 & $1.011 \pm 0.008$  \\ \hline
3 & Cr & $30\pm5$ & $2.02 \pm 0.07$ & 6 & $0.622\pm 0.01$  \\ \hline
4 & Ti & $15\pm3$& $1.08\pm0.02$ & 4 & $0.827 \pm 0.004$   \\ \hline
{5} &  {Ti} &  {$15\pm4$} & $1.25\pm0.03$ & 5 & $0.604\pm 0.006$  \\ \hline
{6} &  {Ti} &  {$29\pm4$} & $1.03 \pm 0.05$ & 4 & $0.805\pm 0.008$  \\ \hline
{7} &  {Ni} & {$26\pm3$} & $0.85 \pm 0.02$ & 5& $0.629 \pm 0.012$ \\ \hline
{8} &  {Ni} & {$29\pm5$}& $0.87 \pm 0.07$ & 4 & $0.621 \pm 0.007$  \\ \hline
{9} &  {Ni}  & {$23\pm3$}& $1.01 \pm 0.10$ & 6 & $0.705 \pm 0.010$    \\ \hline
{10} &  {Ni} &  {$21\pm4$}& $0.88\pm0.04$ & 5& $0.711\pm 0.005$   \\  \hline
\end{tabular}
\caption{\label{muntab} Flake thickness $t$, flake width $W$, number of contacts, and  $d_1$ for each of the measured devices.}
\label{indDev}
\end{table}

\section{Additional Information on Gate Dependence}

\begin{figure}
  \includegraphics[width=\textwidth]{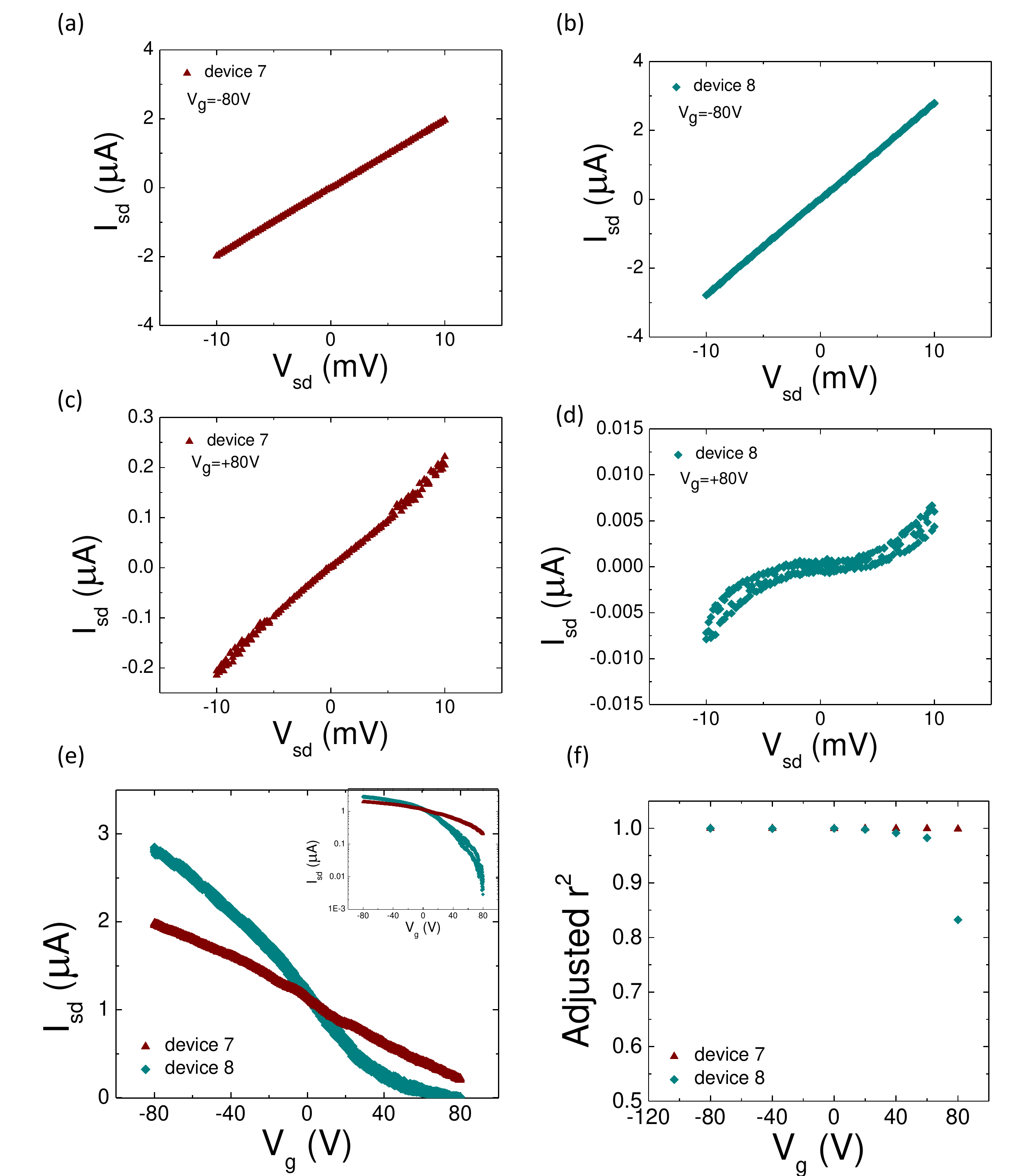}
  \caption{\label{fig:NiVg} Current-voltage characteristics of two different devices with Nickel contacts at back gate voltage (a,b) $+80$~V and (c,d) $-80$~V. Data from device~7 is also shown in Fig.~3 of the main text. (e) Source-drain current as a function of gate voltage for the two devices, which clearly show a difference in modulation. $V_{sd} = 1$~mV. The inset shows the same data on a logarithmic scale. (f) Adjusted $r^2$ of linear fits of all current-voltage measurements acquired at different gate voltage. All measurements at $T = 4.2$~K.}
\end{figure}

To provide further evidence that carrier concentration is a key element to observe a linear trend in the current-voltage traces, we show in Fig.~\ref{fig:NiVg} two devices with contacts from the same metal (Ni), which display a different modulation of the current by gate voltage, device~7 and device~8, as labeled in Table~\ref{indDev}. At negative gate voltage, when the devices are in the accumulation regime, both devices show a linear current-voltage characteristics, see Figs.~\ref{fig:NiVg}(a) and (b), while in the depletion regime (Figs.~\ref{fig:NiVg}(c) and (d)), device~8 has a strongly nonlinear $I-V$ characteristics, with a maximum source-drain current more than one order of magnitude smaller than device~7, which shows a linear $I-V$ characteristics. The $I_{sd}$ vs.~$V_g$ curve, shown in Fig.~\ref{fig:NiVg}(e), clearly shows the differences between these two devices. The on-off ratio, defined as the maximum current obtained in the accumulation regime divided by the minimum current obtained in the depletion regime, is $\approx 10$ for device~7 and $\approx 340$ for device~8. The trend of all $I-V$ curves at different gate voltage is summarized by the values of the adjusted-$r^2$ shown in Fig.~\ref{fig:NiVg}(f). For $V_g > 40$~V, a decrease is observed for device~8. This confirms that reaching the depletion regime results in non-linearities in the current-voltage characteristics.

\pagebreak

\providecommand{\latin}[1]{#1}
\makeatletter
\providecommand{\doi}
  {\begingroup\let\do\@makeother\dospecials
  \catcode`\{=1 \catcode`\}=2 \doi@aux}
\providecommand{\doi@aux}[1]{\endgroup\texttt{#1}}
\makeatother
\providecommand*\mcitethebibliography{\thebibliography}
\csname @ifundefined\endcsname{endmcitethebibliography}
  {\let\endmcitethebibliography\endthebibliography}{}

\end{document}